\documentstyle[emulateapj,epsf]{article}
\submitted{slugcomment}

\def\lesssim{\mathrel{\hbox{\rlap{\hbox{\lower4pt\hbox{$\sim$}}}\hbox{$<$}}}}
\def\gtrsim{\mathrel{\hbox{\rlap{\hbox{\lower4pt\hbox{$\sim$}}}\hbox{$>$}}}}

\def\arcdeg{\hbox{$^\circ$}}

\newcommand{\mm}[1]{\mbox{$#1$}}
\newcommand{\unit}[1]{\ifmmode \:\mbox{\rm #1}\else \mbox{#1}\fi}
\renewcommand{\sb}[1]{\mbox{$_{\rm #1}$}}

\newcommand{\mone}{\mm{^{-1}}}

\newcommand{\kms}{\unit{km~s\mone}}
\newcommand{\kpc}{\unit{kpc}}

\newcommand{\lb}[2]{\mm{l = #1\arcdeg}, \mm{b = #2\arcdeg}}

\newcommand{\dec}{\mm{\delta}}

\renewcommand{\dec}{\mbox{Dec.}}
\newcommand{\secref}[1]{\S\ref{sec:#1}}

\newcommand{\eqref}[1]{equation~(\ref{eq:#1})}

\newcommand{\figref}[1]{Fig.~\ref{fig:#1}}
\newcommand{\tabref}[1]{Table~\ref{tab:#1}}

\newcommand{\old}[1]{}

\newcommand{\mg}{\mbox{Mg$_2$}}
\newcommand{\ebv}{\mbox{$E(\bv)$}}
\newcommand{\bvmg}{{\bv -- \mg}}

\newcommand{\dn}{\mbox{$D\sb{n}$}}

\slugcomment{To Appear PASP, Jan 1999}

\lefthead{M. J. HUDSON} 
\righthead{Systematic Errors in Reddening Maps} 

\begin{document} 
\title{A Test for Large-Scale Systematic Errors in Maps of Galactic
Reddening}

\author{Michael J. Hudson}
\affil{ Department of Physics \& Astronomy, University of Victoria,\\
  P.O. Box 3055, Victoria, B.C. V8W 3P6, Canada\\
  E-mail: hudson@uvastro.phys.uvic.ca}

\begin{abstract} 

Accurate maps of Galactic reddening are important for a number of
applications, such as mapping the peculiar velocity field in the
nearby Universe. Of particular concern are systematic errors which
vary slowly as a function of position on the sky, as these would
induce spurious bulk flow.

We have compared the reddenings of Burstein \& Heiles (BH) and those
of Schlegel, Finkbeiner \& Davis (SFD) to independent estimates of the
reddening, for Galactic latitudes $|b| > 10\arcdeg$.  Our primary
source of Galactic reddening estimates comes from comparing the
difference between the observed \bv\ colors of early-type galaxies,
and the predicted \bv\ color determined from the \bvmg\ relation.  We
have fitted a dipole to the residuals in order to look for large-scale
systematic deviations.

There is marginal evidence for a dipolar residual in the comparison
between the SFD maps and the observed early-type galaxy reddenings.
If this is due to an error in the SFD maps, then it can be corrected
with a small (13\%) multiplicative dipole term.  We argue, however,
that this difference is more likely to be due to a small (0.01 mag.)
systematic error in the measured \bv\ colors of the early-type
galaxies.  This interpretation is supported by a smaller, independent
data set (globular cluster and RR Lyrae stars), which yields a result
inconsistent with the early-type galaxy residual dipole.  BH
reddenings are found to have no significant systematic residuals,
apart from the known problem in the region $230\arcdeg < l <
310\arcdeg$, $-20\arcdeg < b < 20\arcdeg$.
\end{abstract} 


\section{Introduction}

Accurate values of Galactic extinction are essential for a number of
applications, such as measuring distances and peculiar velocities of
galaxies.  A small error of only 0.015 mag.\ in the R-band Galactic
extinction corresponds to a systematic distance error of 1\% for the
Fundamental Plane distance indicator.  When measuring the large-scale
bulk motion of galaxies, the small-scale accuracy of reddening maps is
not critical, since random errors add to the scatter but do not
introduce a systematic bias in the bulk flow.  On the other hand,
large-scale coherent systematic errors in the reddenings would
introduce corresponding systematic errors in the bulk flow.

In this paper, we test two reddenings maps for such large-scale
systematic errors: the reddening maps of Burstein \& Heiles (1982,
hereafter BH), which are based on neutral hydrogen (for $\dec > -23$,
Shane-Wirtanen galaxy counts are used to determine dust-to-gas
ratios); and the maps of Schlegel, Finkbeiner \& Davis (1998,
hereafter SFD) which are based on dust emission measured by IRAS with
dust temperatures determined from DIRBE data.  Note that the two maps
are completely independent.

There is some evidence that the BH reddenings are systematically in
error in at least one large coherent region of the sky.  Burstein et
al.\ (1987, hereafter B87) found that the BH reddenings were
overestimates (by a factor 2) in the region $230\arcdeg < l <
310\arcdeg$, $-20\arcdeg < b < 20\arcdeg$ (hereafter referred to as
the Vela region). Note that this region lacks a dust-to-gas ratio
estimate.

SFD calibrated their reddening maps using the colors of brightest
cluster galaxies from Postman \& Lauer (1995) and tested their maps
with the \bvmg\ relation and the data of Faber et al.\
(1989). However, SFD did not test for coherent residuals as a function
of longitude.  Such residuals are the bane of peculiar velocity work;
they are the focus of this paper.

To test the BH and SFD maps, we require a sample of distant objects
with standard colors.  Our primary sample is the early-type galaxy
data set of Faber et al.\ (1989).  We will make use of the tight
correlation between \mg\ index and \bv\ color.  Lynden-Bell et al.\
(1988) used the \bvmg\ relation to show that putative systematic
errors in the BH reddenings could not be large enough to explain the
bulk flow of $570 \kms$ towards \lb{307}{9}.  However, Lynden-Bell et
al.\ did not attempt to push this farther and obtain constraints on
the accuracy of the BH reddenings. This was perhaps due to systematic
uncertainties in the \mg\ index, which are at the level of $\sim 0.01$
in \mg\ (Davies et al.\ 1987), equivalent to 0.008 in \ebv\ or 0.03 in
$B$-band extinction.

We have recently determined the corrections required to bring \mg\
indices from 23 different spectroscopic systems onto a standard
system, by comparing 1784 repeat \mg\ measurements for 418 galaxies
(Smith et al.\ 1997, Hudson et al.\ 1999).  As a result the
uncertainties in the corrections to the 7S \mg\ systems have been
reduced by a factor 5.  This allows us to re-examine the reddening
maps with increased accuracy.

There has been much discussion in the literature of the value of
reddening at the poles, i.e.\ the zero-point of the reddening maps.
This zero-point is impossible to calibrate using only extragalactic
color standards.  Fortunately global zero-point errors have no effect
on peculiar velocity work since they are absorbed into the zero-point
of the distance indicator.  We make no attempt in this paper to
determine the correct zero-point of the reddening maps.

The outline of this paper is as follows.  In \secref{ellip}, we use
early-type galaxies as reddening estimators, and present our data and
method (\secref{method}), results (\secref{results}) and discuss the
effect of photometric errors in the \bv\ data (\secref{discuss}). In
\secref{gcrr}, we constrain the systematic errors in the SFD maps
using independent reddening estimates obtained from globular clusters
and RR Lyrae stars, and we summarize our results in \secref{summary}.

\section{Reddening from Early-Type Galaxies}
\label{sec:ellip}
\subsection{Data and Method}
\label{sec:method}

Our primary data are the early-type galaxies of the `Seven Samurai'
(7S) survey (Faber et al.\ 1989), with \bv\ colors measured within 67
arcsec from B87.  However, we use revised \mg\ data for these galaxies
which are corrected for offsets between \mg\ measurements obtained
from different telescopes (Hudson et al.\ 1999). Our corrected \mg\
values agree well with the subset of the 7S data recently published by
Trager et al. (1998). We use only the \bv\ data with \dn\ photometric
quality of 1, corresponding to random and systematic errors less than
0.02 dex (B87)%
. The data are given in \tabref{ellip}.
Only the first and last four lines of the data are reproduced in the
printed version. The full table is accessible electronically.
Finally, we limit our analyses to Galactic latitudes $|b| > 10$.
The \bvmg\ relation has a scatter of only 0.028 mag.  This relation is
shown in \figref{bvmg}.
\vbox{%
\begin{center}
\leavevmode
\hbox{%
\epsfxsize=8.9cm
\epsffile{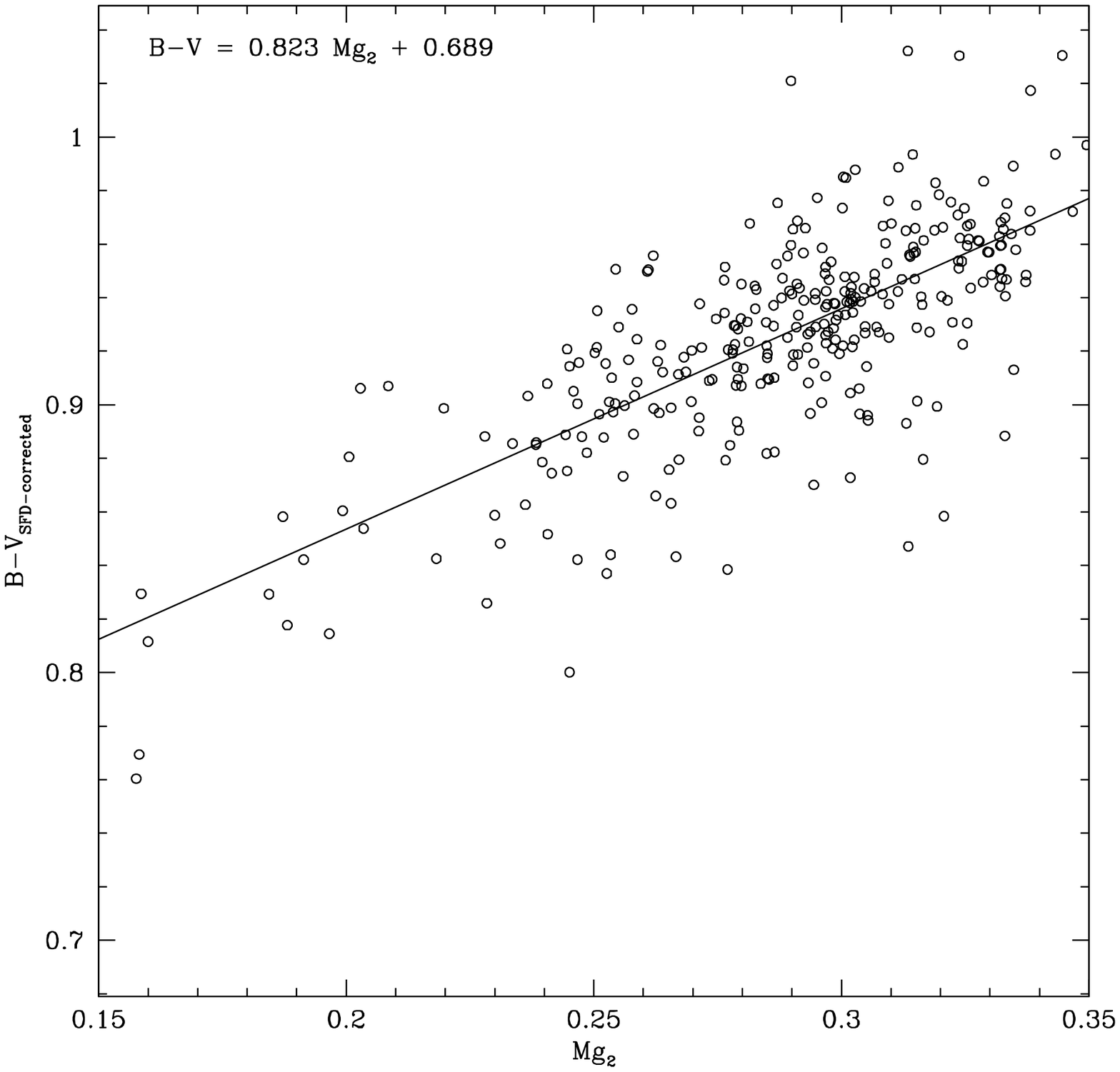}}
\begin{small}
\figcaption{%
The \bvmg\ relation with \bv\ corrected using SFD
reddenings. Four galaxies with $\mg < 0.15$ are not shown.  
\label{fig:bvmg}}
\end{small}
\end{center}}

\begin{table*}
\caption{Early-type galaxies \label{tab:ellip}}
\begin{tabular}{lrrrrrrrrr}
\tableline
\multicolumn{1}{c}{Name} & 
\multicolumn{1}{c}{\bv} & 
\multicolumn{1}{c}{$l$} & 
\multicolumn{1}{c}{$b$} & 
\multicolumn{1}{c}{\mg} & 
\multicolumn{1}{c}{Qual} & 
\multicolumn{1}{c}{\ebv$_{\rm SFD}$} & 
\multicolumn{1}{c}{\ebv$_{\rm BH}$} & 
\multicolumn{1}{c}{Clus\tablenotemark{a}} & 
\multicolumn{1}{c}{$\sigma_{cl}$\tablenotemark{b}} \\
 &  &  &  &  &  &  &  &  & \kms \\
\tableline \\
A1656:D-107 & 0.87 &  56.7953 &  88.0890 & 0.2311 & 111 & 0.0118 &  0.0114 & 1 &  821 \\
A1656:D-125 & 0.92 &  56.9377 &  87.9992 & 0.2531 & 111 & 0.0089 &  0.0121 & 1 &  821 \\
A1656:D-136 & 0.90 &  60.1584 &  88.1393 & 0.2672 & 111 & 0.0105 &  0.0120 & 1 &  821 \\
A1656:D-153 & 0.95 &  58.7005 &  87.9678 & 0.2747 & 111 & 0.0079 &  0.0124 & 1 &  821 \\
\multicolumn{10}{c}{$\cdots$}\\
N7626       & 1.04 &  87.8581 & -48.3781 & 0.3260 & 111 & 0.0725 &  0.0411 & 1 &  212 \\
N7768       & 0.94 & 106.7153 & -33.8139 & 0.3153 & 212 & 0.0386 &  0.0276 & 1 & -999 \\
N7785       & 1.00 &  98.5380 & -54.2775 & 0.2915 & 212 & 0.0564 &  0.0420 & 0 &    0 \\
U03696      & 1.01 & 168.5668 &  22.8446 & 0.3011 & 312 & 0.0715 &  0.0716 & 1 &  327 \\
 &  &  &  &  &  &  &  &  & \\
\tableline
\end{tabular}
\tablenotetext{a}{Flag for cluster/group (set to 1) or field (0) }
\tablenotetext{b}{Cluster/group velocity dispersion ($-999 =$ no data)}
\end{table*}

If we compare the \bvmg\ relations of the subsample of cluster
galaxies to the subsample of galaxies in groups (with velocity
dispersions [Hudson 1993] less than $\sim 400$ \kms) or in the field,
we find that the cluster galaxies are redder than group/field galaxies
by $0.01\pm0.004$ mag at a given \mg.
Hudson (1999) shows that Malmquist bias effects can cause offsets
between field and cluster samples at this level, even when
distance-independent measures (\bv\ and \mg) are compared. For the
purposes of this paper, however, we are not concerned with the
interpretation of this offset.  We simply correct the \bv\ colors of
field galaxies by adding 0.01.  The conclusions of this paper,
however, are not affected if we neglect this correction.

\begin{figure*}
\figurenum{2}
\plotone{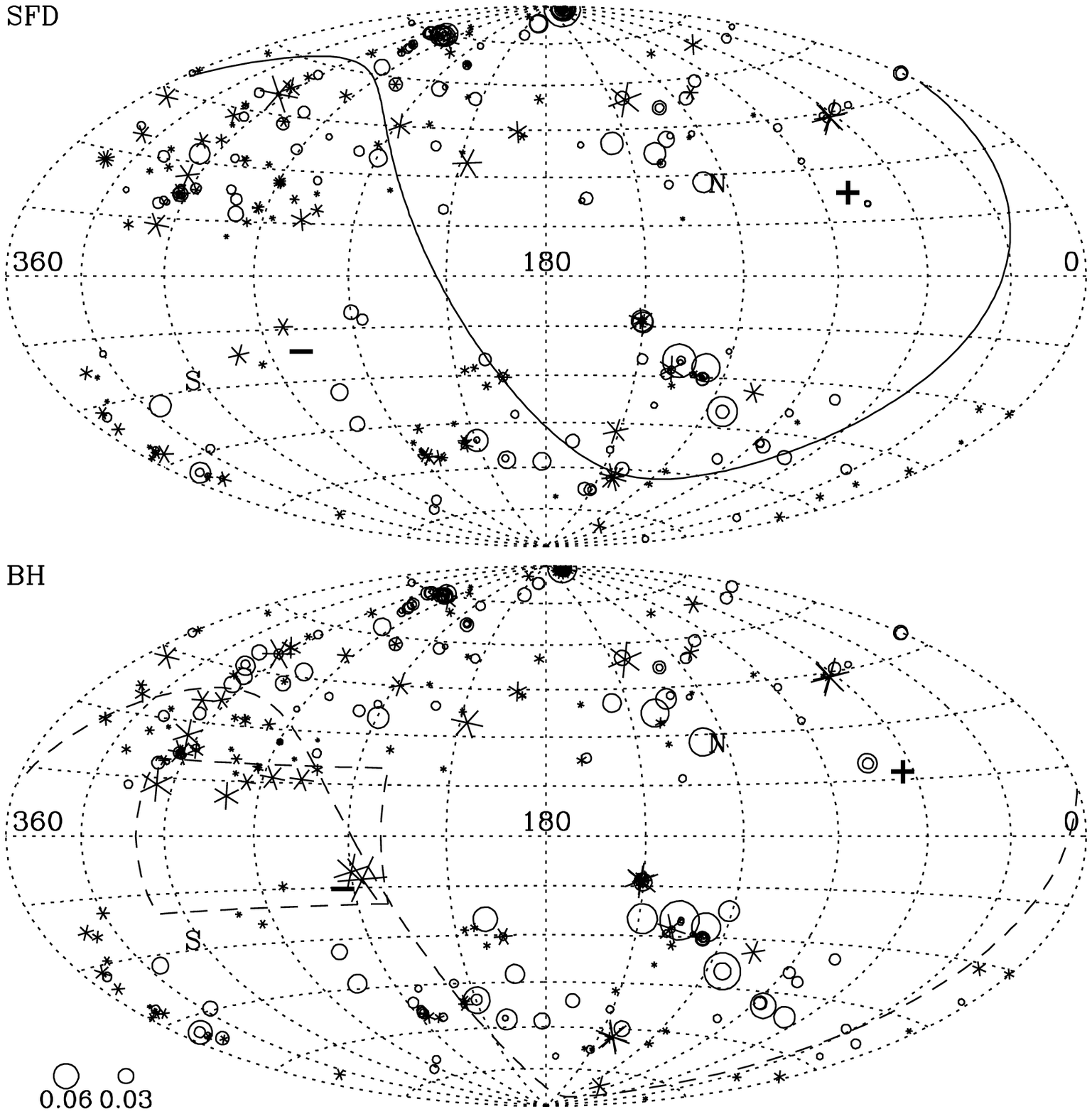}
\caption{%
Early-type galaxy residuals \ebv\sb{ellip} - \ebv\sb{map} are shown on
the sky in Aitoff projection in Galactic coordinates.  Open circles
show galaxies with positive residuals, i.e. galaxies for which the
maps underpredict the reddening; crosses show galaxies with negative
residuals.  The size of the symbol scales with the absolute value of
the residual. The symbols in the lower left corner show the sizes of
symbols corresponding to residuals of 0.06 and 0.03 mag. The top panel
shows the residuals from the SFD maps; the poles of the SFD correction
are marked by the plus and minus symbols. Celestial North and South
poles are indicated by 'N' and 'S'.  The solid line is the
equator. The lower panel shows the BH residuals. The dashed line is at
${\rm Dec.}  = -23\arcdeg$ and divides the region with galaxy counts
from the region without.  The dashed box shows the Vela region.
\label{fig:aitoff}
}
\end{figure*}

The residuals from the \bvmg\ relation for both SFD and BH reddenings
are plotted on the sky in an Aitoff projection in \figref{aitoff}.  No
obvious trend to the residuals is immediately apparent although the
Vela region stands out in the BH plot, and to a lesser extent in the
SFD plot.  Clearly any deviation of the residuals from isotropy is at
a fairly low level.  To measure the anisotropy requires a specific
model.

Our method is straightforward: we use the tight intrinsic relation to
predict the true \bv\ and compare this with the observed
$\bv\sb{obs}$, to obtain a reddening estimate
\begin{equation}
E(\bv)\sb{Ellip} = (\bv)\sb{obs} - [A + B \, \mg]
\end{equation}
where $A$ and $B$ are free parameters.  Since our goal is to look for
coherent residuals from the SFD and BH maps, we fit the map reddening
to these reddening estimates allowing for a correction term.  For the
form of this correction, we adopt the lowest order spherical harmonic,
i.e.\ a dipole.  We have considered both additive and multiplicative
dipole corrections to the reddenings.  We adopt a multiplicative
correction, which is perhaps more realistic (e.g. a coherent error in
the dust temperature).  The reddening estimates are then fit by
\begin{equation}
E(\bv)\sb{Ellip} = (1 + F \cdot \hat{r}) \, \ebv\sb{map}
\end{equation}
where $\ebv\sb{map}$ is the predicted reddening given by BH or SFD.
$F$ is a dipole vector and $\hat{r}$ is the unit vector towards the
galaxy.  In practice, these two steps are performed simultaneously.

The residuals around these relations are slightly non-Gaussian, with a
tail of blue objects. In order that our fit be robust, we minimize the
sum of absolute deviations of the residuals. This is equivalent to
fitting the median (Press et al.\ 1992).  We estimate the significance
of the dipole coefficients by performing Monte Carlo simulations in
which the positions of early-type galaxies are randomized while their
\bv\ and \mg\ values are kept fixed.

\subsection{Results}
\label{sec:results}

\begin{table*}
\caption{Dipoles of \ebv\ residuals \label{tab:dipoles}}
\begin{tabular}{lrrlrrrrr}
\tableline
\multicolumn{1}{c}{Sample} & 
\multicolumn{1}{c}{$N_{\rm obj}$} & 
\multicolumn{1}{c}{scatter} & 
\multicolumn{1}{c}{Map} & 
\multicolumn{1}{c}{Corr\tablenotemark{a}} & 
\multicolumn{1}{c}{$|F|$} & 
\multicolumn{1}{c}{$l$} & 
\multicolumn{1}{c}{$b$} & 
\multicolumn{1}{c}{Prob\tablenotemark{b}} \\
\tableline
& & & & & &  &  & \\
Ellip  & 311 & 0.028 & SFD & 0.000 & 0.13 &  80 &  23 &  8.4 \\
Ellip\tablenotemark{c}  & 296 & 0.031 & BH  & 0.000 & 0.17 &  64 & 17 & 44.7 \\
Ellip\tablenotemark{c}  & 296 & 0.028 & SFD  & 0.000 & 0.13 & 94 & 15 & 13.1 \\
Ellip  & 311 & 0.028 & SFD & 0.010 & 0.07 & 328 & -69 & 57.2 \\
Ellip\tablenotemark{c}  & 296 & 0.031 & BH  & 0.010 & 0.11 & 47 & -27 & 87.7 \\
GC+RR  & 137 & 0.028 & SFD & 0.000 & 0.08 & 312 & -47 & 95.0 \\
Ellip+GC+RR & 448 & 0.028 & SFD & 0.000 & 0.05 & 98 & -35 & 38.6 \\
\tableline
\end{tabular}

\tablenotetext{a}{Correction applied to Northern \ebv\ data}
\tablenotetext{b}{Probability that measured dipole could arise by chance}
\tablenotetext{c}{Excluding the Vela region}
\end{table*}

The result of the dipole fits to both SFD and BH are given in
\tabref{dipoles}.  For the SFD \mg\ fit, we detect a marginally
significant dipole (at the 92\% CL) of $|F\sb{SFD}| =
0.13_{-0.10}^{+0.06}$ towards \lb{80}{23} (errors correspond to 66\%
confidence range).  Note that the anti-apex of this dipole is
\lb{240}{-23}, which is in the Vela region.

We detect no significant residual dipole in the BH reddenings, {\em
if\/} the Vela region is excluded from the analysis. (If we exclude
this region from the SFD analysis, the significance of the SFD dipole
term drops to 87\% CL).  The lower significance of the BH dipole
compared to SFD should not be misinterpreted as evidence that its
residual dipole is lower -- it is not -- rather the reduced
significance is due to the larger scatter in the residuals when BH
reddenings are used, which leads to a larger error in the dipole:
$|F\sb{BH}| = 0.17_{-0.17}^{+0.13}$.

\begin{figure*}
\figurenum{3}
\plotone{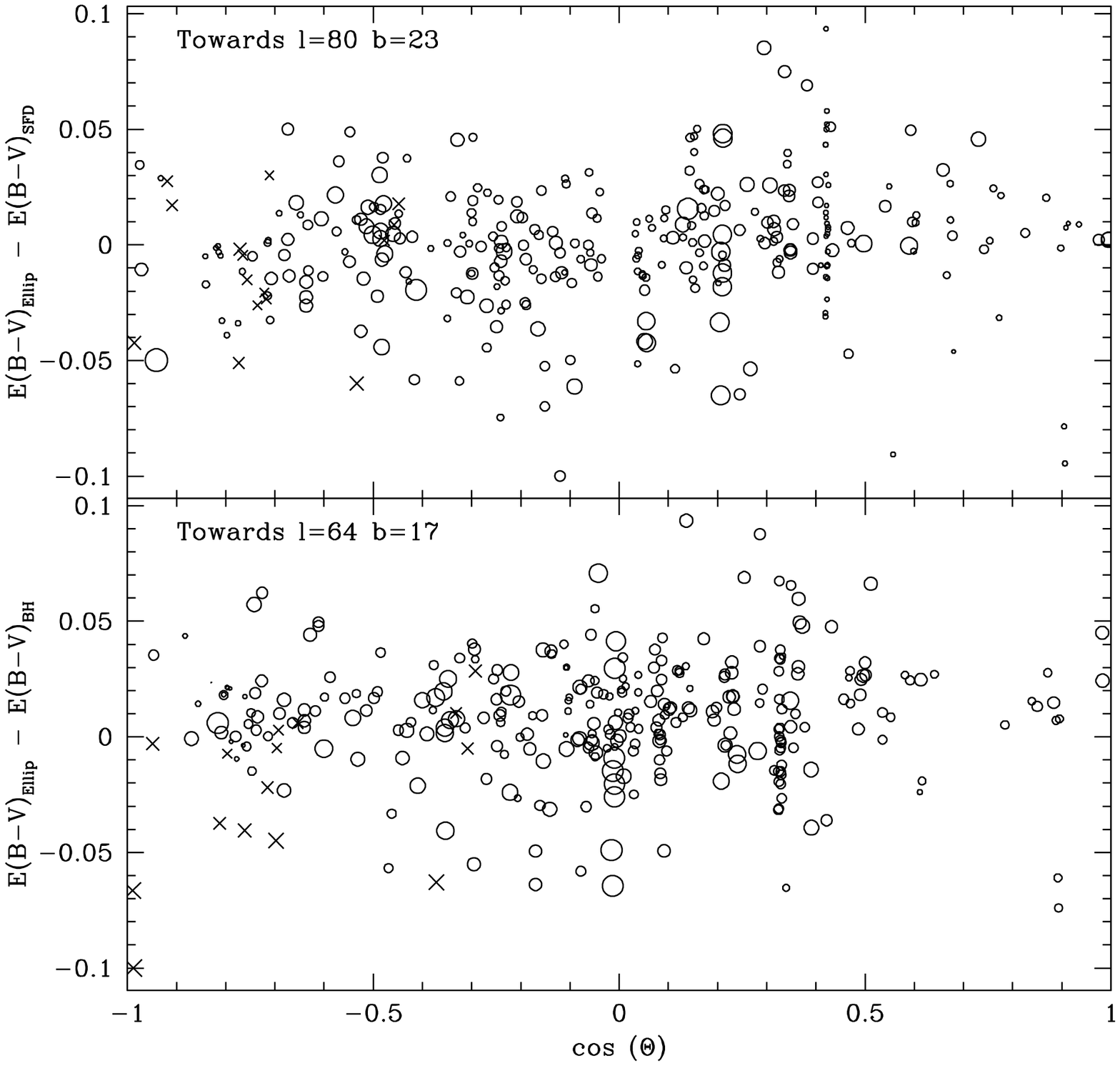}
\caption{%
Early-type galaxy extinction residuals versus cosine from dipole direction.  
The symbol area is proportional to \ebv\sb{map}. Crosses indicate
galaxies in the Vela region; open circles for all other galaxies.
Top panel: SFD \ebv. Bottom panel: BH \ebv.
\label{fig:dipole}
}
\end{figure*}

\figref{dipole} shows the residuals from the \bvmg\ relation as a
function of the cosine of the angle from the apex (\lb{80}{23} for SFD
and \lb{64}{17} for BH).

\subsection{Discussion}
\label{sec:discuss}
We have modeled the residuals with a dipole term.  However, it is also
possible that the residuals are clustered in a few, smaller-scale
(30-60\arcdeg) patches on the sky.  We have already discussed the Vela
region, in which both the BH reddenings as well as those of SFD
overestimate the reddening.  The region $120\arcdeg < l < 180\arcdeg,
0\arcdeg < b < 40\arcdeg$, (hereafter referred to as the Cameleopardis
region) also stands out in \figref{aitoff} due to its excess of red
galaxies.  The Cameleopardis region is opposite on the sky to Vela
region, so it is possible that these two regions together are the
source of the dipole signal.  The sparseness of the data do not allow
us to determine whether the residuals are confined to smaller areas,
or whether they vary more slowly (like a dipole).

At low Galactic latitudes, systematic errors are expected to be
largest, both in the predictions and in the observed colors, due to
the problems of contamination by foreground stars in the Milky Way.  If
we limit the analysis to Galactic latitudes $|b| > 20\arcdeg$,
however, we still find an identical dipole for SFD and a somewhat
larger dipole for BH, compared to the results found above for $|b| >
10\arcdeg$.

So far we have interpreted these residuals as errors in the
reddenings.  It is possible that systematic errors in the \mg\ data or
\bv\ data are responsible for these residuals.  The \mg\ data for the
7 spectroscopic systems tabulated by 7S were corrected for systematic
offsets, but the uncertainty in these corrections was 0.01
mag. (Davies et al.\ 1987). By inter-comparison of common data, these
\mg\ data have been re-corrected and brought onto a common system
(Hudson et al.\ 1999).  While the corrections are large (as much as
0.02), the uncertainties in these corrections are now considerably
smaller (typically 0.0025 mag).  This translates to only 0.002 mag in
\ebv\ which is a factor of 5 smaller than the residual dipole.
Therefore systematic errors in \mg\ cannot account for the residuals.

It is also possible that there are systematic errors in the \bv\
colors.  We have performed fits in which we allow for a systematic
error in color between galaxies in the northern and southern celestial
hemispheres.  In \tabref{dipoles} we also show fits for BH and SFD
after a 0.01 mag correction applied to the Northern \bv\ values.  When
this correction is applied, the residual dipole terms are small and
consistent with zero.  Thus an offset in photometry can account for
the residuals in the SFD comparison.  In fact, the significance of
this photometric correction is somewhat higher than the dipole term
fitted above.

As discussed above, we cannot really distinguish between photometric
errors in small patches of the sky and large-scale (north-south)
photometric offsets.  The \bv\ colors tabulated in B87 arise from 10
different photometric systems. Table 4B of B87 indicates that, while
most of their photometric systems are of high accuracy (better than
0.01 mag. in \bv), there are several exceptions. Most notable of these
is the K3 system (KPNO 0.9m Oct 81)%
\footnote{During the K3 run a can of oil was found after the end of
the run and an oil slick covered part of the mirror (B87).}.  The K3
system shows a systematic offset of 0.02 mag in \bv\ and possibly a
weak (uncorrected) color term (B87).  Note that about half of the
galaxies in the Cameleopardis region have \bv\ colors from the K3
system, so this may account for some of the deviant galaxies in this
region.  It is beyond the scope of this paper to extract the K3
observations from the merged photometry of B87.  We simply note that
although one can interpret the residuals as a north-south photometric
offset, it is also possible that errors from one or more subsystems
could mimic the same effect.

Note that in \figref{aitoff} both BH and SFD have residuals in the
same locations (e.g.\ in the Vela and Cameleopardis regions).  Since
the two reddening maps are independent, this argues that the primary
source of these residuals is due to photometric errors.

Therefore, while the residual dipole in the SFD comparison may be
real, we conclude that it is more likely due to photometric errors.
An independent sample is required to test this residual dipole more
thoroughly.

\section{Reddening Estimates from Globular Clusters and RR Lyrae stars}
\label{sec:gcrr}

We can perform an independent test of the reddenings using the
estimated \ebv\ toward distant globular clusters (GCs) and RR Lyrae
stars.

We take GC \ebv\ estimates from the compilation of Harris (1996)%
\footnote{Available electronically at
http://www.physics.mcmaster.ca/GC/mwgc.dat.}%
, but exclude clusters
with $|b| < 10$ and distance perpendicular to the Galactic Plane, $|Z|
< 3$ \kpc.  This leaves 50 GCs.  We find that, in the mean, the
estimated reddenings of the GCs are lower than those of SFD by $-0.008
\pm 0.004$ mag.  This paper is not concerned with zero-point errors
(the monopole term) because errors of this type drop out of many
applications such as peculiar velocity studies.  We have therefore
simply increased all globular cluster \ebv\ by 0.01 mag.  The scatter
of \ebv\ is 0.027 mag.

\begin{table*}
\caption{RR Lyrae and Abt \& Golson stars \label{tab:rr}}
\begin{tabular}{rrrr}
\tableline \\
\multicolumn{1}{c}{$l$} & 
\multicolumn{1}{c}{$b$} & 
\multicolumn{1}{c}{\ebv} & 
\multicolumn{1}{c}{\ebv$_{\rm SFD}$} \\
\tableline \\
115.70 & -33.10 &  0.058 & 0.044 \\
128.40 & -23.60 &  0.035 & 0.047 \\
 57.90 & -34.00 &  0.006 & 0.052 \\
 53.20 & -44.30 &  0.074 & 0.048 \\
\multicolumn{4}{c}{$\cdots$}\\
304.60 &  57.40 & -0.022 & 0.030 \\
323.40 &  67.50 & -0.020 & 0.035 \\
123.90 &  18.60 &  0.249 & 0.247 \\
122.00 &  22.10 &  0.112 & 0.138 \\
\tableline
\end{tabular}
\end{table*}

\begin{figure*}
\figurenum{4}
\plotone{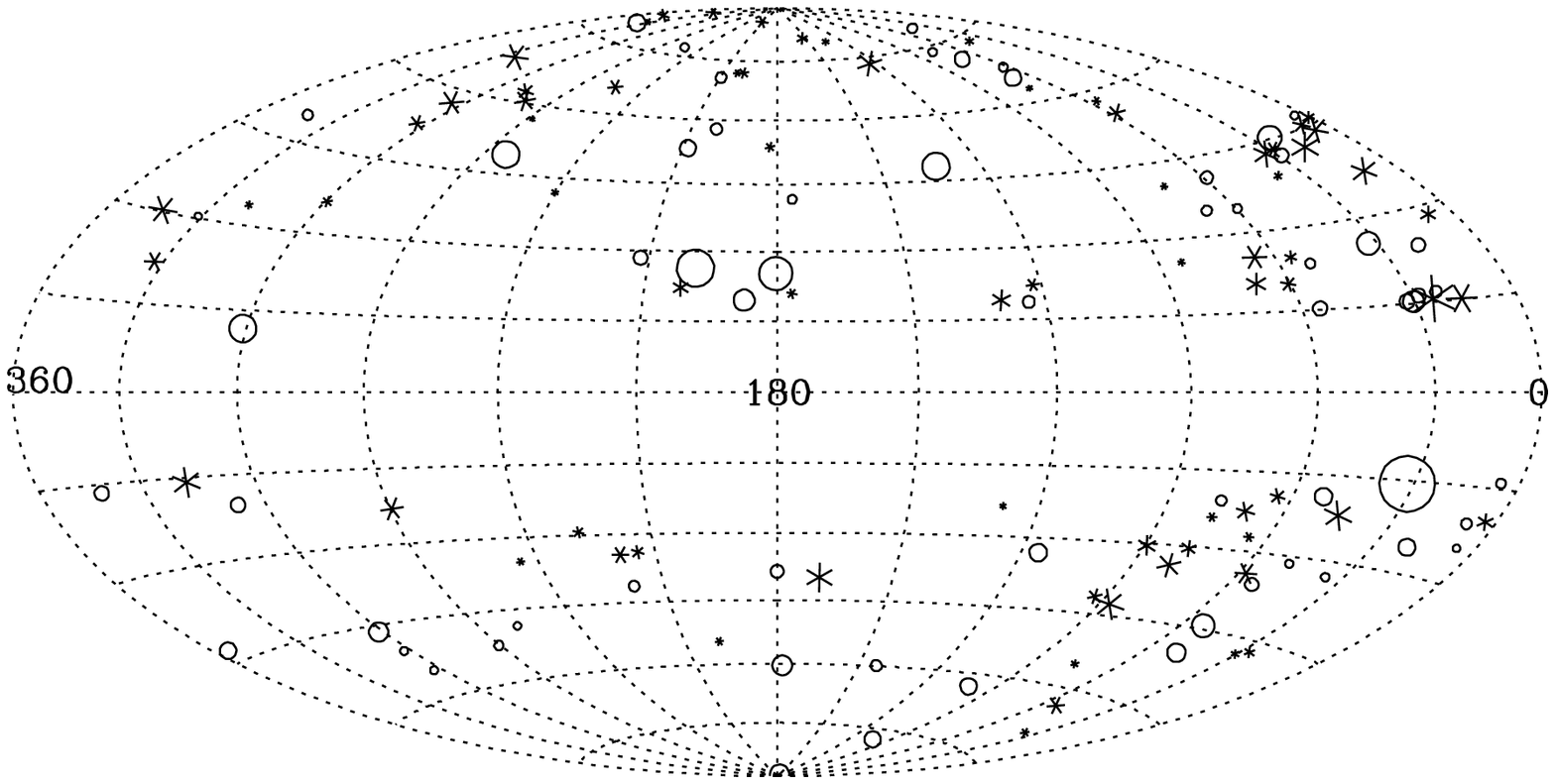}
\caption{%
Globular cluster and RR Lyrae residuals $\ebv\sb{GC/RR} -
\ebv\sb{SFD}$ on the sky in an Aitoff projection of Galactic
coordinates.  Symbols are as in \figref{aitoff}.
\label{fig:gcrr}
}
\end{figure*}

We have also use the 86 RR Lyrae stars from the sample of Burstein \&
Heiles (1978).
The data are reproduced in \tabref{rr}.
Only the first and last four lines of the data are given in the
printed version. The full table is accessible electronically.
We find, in the mean, that the RR Lyrae sample has \ebv\ lower than
the SFD predictions by $-0.016\pm0.003$ mag.  We have consequently
increased all RR Lyrae \ebv\ by 0.01 mag.  The scatter around the SFD
values is 0.027 mag, the same as found for the GCs and for the
residuals from the early-type galaxy \bvmg\ relation.  The residuals
for these two samples are shown in \figref{gcrr}.

Both the GC and RR Lyrae samples are sufficiently small that they do
not place strong constraints on the residual dipole.  Taken together,
the best fit dipole is $F = 0.08\pm0.07$ towards \lb{312}{-47},
although this result is not statistically significant; the GC/RR Lyrae
sample is consistent with no residual dipole.  Note, however, that
this GC/RR $F$ is in the opposite sense to the SFD $F$ found above. We
can thus use the combined GC and RR Lyrae sample as an independent
test of the early-type \bvmg\ residual dipole.  We generate 1000
bootstrap samples of the globular cluster and RR Lyrae solution and
project these along \lb{80}{23}, the \bvmg\ dipole direction.  We
obtain a residual as large as the \bvmg\ case only 2\% of the time.
This argues that the \bvmg\ dipole found above is more likely to be
due to a systematic photometric offset between north and south.  In
any case, it establishes the \bvmg\ dipole result as the upper limit
to the large-scale errors in the SFD maps.

\section{Summary}
\label{sec:summary}

We have compared the SFD and BH reddenings to an independent reddening
estimate, namely the difference between the observed \bv\ colors of
early-type galaxies, and the predicted \bv\ color determined from the
\bvmg\ relation.  Specifically, we fit a dipole to the residuals in
order to look for large-scale systematic deviations.  
For BH, we find no significant evidence for a dipole once the Vela
region is excluded.  For SFD, we find marginal (92\% CL) evidence of a
dipole in the \ebv\ residuals.  We are unable to distinguish between
two possibilities: (1) a small multiplicative dipolar correction term
to the SFD reddenings (a 13\% increase towards \lb{80}{23}, and a
corresponding decrease on the other side of the sky), or (2) a small
photometric error of 0.01 mag in \ebv\ between northern and southern
hemispheres.  We argue that the latter option is more likely, for the
following reasons: (1) 0.01 mag.\ is well within the expected
photometric errors, and (2) the dipole for the BH reddenings is very
similar to that of SFD in both amplitude and direction, even though
the BH and SFD maps are independent.  The sparseness of the data do
not allow us to determine whether the residuals are actually due to
smoothly-varying large-scale (dipolar) errors in SFD reddenings or in
the photometry, or whether the problem lies in specific regions, e.g.\
the Vela region and/or the Cameleopardis region.

The sparser globular cluster and RR Lyrae reddening data are
consistent with no dipole residual. Nevertheless, from this data set,
we can exclude the dipole obtained from the early-type galaxies at the
98\% confidence level.

Taken together, these results suggest that there are no large-scale
errors in the SFD maps, or in the BH maps (outside the Vela region).
The 13\% dipole correction to SFD should be taken as a reasonable
upper limit on any large-scale error in those maps.

\end{document}